\documentstyle{lamuphys}

\begin{document}

\title{Arbitrary spin massless bosonic
fields in d-dimensional anti-de Sitter space}

\author{R.\,R.\,Metsaev}

\institute{Department of Theoretical Physics, P.N. Lebedev Physical
Institute, Leninsky \\ prospect 53, 117924, Moscow, Russia}

\maketitle

\begin{abstract}
Arbitrary spin free massless bosonic fields
propagating in even $d$ - dimensional anti-de Sitter spacetime are
investigated. Free wave equations of motion, subsidiary
conditions and the corresponding gauge transformations for such
fields are proposed.  The lowest eigenvalues of the energy
operator for the massless fields and the gauge parameter fields
are derived.  The results are formulated in $SO(d-1,2)$
covariant form as well as in terms of intrinsic
coordinates. An inter-relation of two definitions of masslessness
based on gauge invariance and conformal invariance is discussed.
\footnote{Talk given at Dubna International Seminar
`Supersymmetries and Quantum Summetries'  dedicated to
the memory of Victor I. Ogievetsky, Dubna, 22-26 June, 1997;
hep-th/9810231}
\end{abstract}

{\it Motivation}.
Some time ago a completely self-consistent interacting equations
of motion for higher massless fields of all spins have been
discovered (\cite{Vas1}). First these equations have been
formulated for the case of four dimensional $d=4$ AdS
spacetime.  Then because the equations allow very natural
generalization to higher spacetime dimensions $d>4$ they have
immediately been extended to such the dimensions (\cite{Vas2}).
These equations are formulated in terms of
wavefunctions $\Psi(x,Z)$ which depend on usual spacetime
coordinates $x^\mu$ and certain twistor like variables
$Z^\alpha$. Usual physical fields as well as certain auxiliary
fields are obtainable by expanding $\Psi(x,Z)$ in powers of $Z$:
$
\Psi(x,Z)=\sum_0^\infty Z^{\alpha_1}\ldots Z^{\alpha_n}
\Phi(x)_{\alpha_1\ldots \alpha_n}$,
where
$\Phi=\{\Phi_{phys},\Phi_{aux}\}$.
For the case of $4d$ theories it is well-established (\cite{Vas2}) that
$\Phi_{phys}$ satisfy the equations of motion which at free
level are equivalent to those investigated in
(\cite{F2}). As to $d>4$ theories, although such a statement is
not proved, it is believed that equations of motion suggested
also describe massless higher spin fields in a
self-consistent way. Unfortunately in contrast to the
completeness of description for $d=4$ little was known  about
the higher spin massless spin fields in arbitrary $d>4$ even
at the level of free fields, unless considerations are
restricted to totally symmetric tensor (or tensor-spinor) fields
(\cite{Vas3},\cite{Vas4}). Filling this gap was a motivation of
our investigation (\cite{Metsit1}-\cite{Metsit4}).
Here we report summary of results.

{\it Setting up the problem}.
First, let us remind the main fact about representations of
anti-de Sitter algebra $so(d-1,2)$
$$ [J^{AB},J^{CD}]
=\eta^{BC}J^{AD}\pm 3 \hbox{ terms}\,,
\qquad
\eta^{AB}=(-,-,+,\ldots,+)\,,
$$
$A,B=0^\prime,0,1,\ldots, d-1$, which are relevant for
elementary particles.
A positive-energy lowest weight irreps of $so(d-1,2)$ algebra
denoted as $D(E_0,{\bf h})$, are defined by $E_0$ which is
lowest eigenvalue of energy operator ${\rm i}J^{00^\prime}$
and by
\begin{equation}\label{mineq1}
{\bf h}=(h_1,\ldots,h_\nu)\,,
\qquad
h_1\geq \ldots \geq h_\nu \geq 0\,,
\qquad
\nu\equiv (d-2)/2
\end{equation}
which is the highest weight of representation of the $so(d-1)$
algebra.  Since in the representations under considerations the
energy is by definition bounded from below the $D(E_0,{\bf h})$
contains the vacuum $|\phi^{E_0}({\bf h})\rangle$
annihilated by all those elements of $so(d-1,2)$ which decrease
the energy. This vacuum forms a linear space which is invariant
under the action of the energy operator ${\rm i}J^{00^\prime}$
and elements of the $so(d-1)$ algebra. In other words
$|\phi^{E_0}({\bf h})\rangle$ is (a) the eigenvalue vector of
${\rm i}J^{00^\prime}$; (b) a weight ${\bf h}$ representation of
the $so(d-1)$ algebra. To expose these properties of
$|\phi^{E_0}({\bf h})\rangle$ it is convenient a
usage of the coordinates ($z$,$\bar{z}$,$y^I$), $I=1,\ldots d-1$,
where
$z=(y^{0^\prime}+{\rm i} y^0)/\sqrt{2}$,
$\bar{z}=z^*$ and the $\eta^{AB}$ takes the nonvanishing
elements $\eta^{z\bar{z}}=-1$, $\eta^{IK}=\delta^{IK}$. In these
coordinates the generators $J^{AB}$ split into
$J^{z\bar{z}}$-energy operator, $J^{zI}$--spin deboost
operator, $J^{\bar{z}I}$--spin boost operator and
$J^{IK}$- generators of the $SO(d-1)$ group.
Now the $|\phi^{E_0}({\bf h})\rangle$ is defined by the
relations
$$
J^{z\bar{z}}|\phi^{E_0}({\bf h})\rangle
=E_0|\phi^{E_0}({\bf h})\rangle\,,
\qquad
J^{zI}|\phi^{E_0}({\bf h})\rangle=0\,.
$$
Then the representation space $D(E_0,{\bf h})$  can be built
by acting with boost operator $J^{\bar{z}I}$ on the vacuum
$|\phi^{E_0}({\bf h})\rangle$:
$$
D(E_0,{\bf h}) =
\sum_{n=0}^\infty \oplus
J^{\bar{z}I_1}\ldots J^{\bar{z}I_n}
|\phi^{E_0}({\bf h})\rangle\,.
$$
It turns out that for certain value of $E_0$  there is singular
vector on the first energy level. If we factorise whole
space by space built on this singular vector then we get
irreducible representations which is, by definition, massless
representation. Now the problem solution to which we are going
to provide can be formulated as follows. Find (i) $E_0$
corresponding to massless representation; (ii) second order
relativistic equations of motion whose space of solution is a
carrier for massless representations; (iii) corresponding gauge
transformations. The relevant $E_0$ for $d=4$
and corresponding field theoretical realization has
been found in (\cite{evans}) and (\cite{F2}) respectively
(for review see \cite{Nic}).

{\it Summary of results}.
We describe the AdS spacetime as a hyperboloid
\begin{equation}\label{hyperbol}
\eta_{AB}^{} y^A y^B=-1\,,
\end{equation}
in $d+1$- dimensional pseudo-Euclidean space with metric tensor
$\eta_{AB}^{}$. The indices $A,B$ are raised and lowered by
$\eta^{AB}$ and $\eta_{AB}^{}$ respectively.  In what follows to
simplify our expressions we will drop the metric tensor
$\eta_{AB}^{}$ in
scalar products.  As is usual, we split the generators $J^{AB}$
into an orbital part $L^{AB}$ and a spin part $M^{AB}$:
$J^{AB}=L^{AB}+M^{AB}$.
The realization of $L^{AB}$ in terms of differential operators
defined on the hyperboloid (\ref{hyperbol}) is:
$$ L^{AB}=y^A
\nabla^B-y^B \nabla^A\,, \qquad
\nabla^A\equiv\theta^{AB}\frac{\partial}{\partial y^B}\,,\qquad
\theta^{AB}\equiv \eta^{AB}+y^Ay^B\,.
$$
The tangent derivative $\nabla^A$ satisfies relations
$$
[\nabla^A,y^B]=\theta^{AB}\,,
\qquad
[\nabla^A,\nabla^B]=- L^{AB}\,,
\qquad
y^A\nabla^A=0\,,\qquad \nabla^A y^A=d\,.
$$
A form for $M^{AB}$ depends on the
realization of the representation. We will use the tensor
realization of representation. As the carrier for
$D(E_0,{\bf h})$ we use of tensor field of the $SO(d-1,2)$ group
\begin{equation}\label{generic}
A^{C({\bf h})}=
A_{\phantom{1}}^{C^1_1,\ldots, C^{h_1}_1,\ldots,
C^1_\nu,\ldots, C^{h_{_\nu}}_\nu}
\end{equation}
defined on the hyperboloid  (\ref{hyperbol}).
By definition, $A^{C({\bf h})}$ is a tensor field whose
$SO(d-1,2)$ indices $C({\bf h})$
have the structure of the Young
tableaux ($YT$) corresponding to the irreps of the $SO(d-1,2)$
group labeled by ${\bf h}$. In what
follows we use the notation ${\sf YT}({\bf h})$ to indicate
such $YT$. The $h_i$ indicates the number of boxes in the $i$-th
row of ${\sf YT}({\bf h})$.  To simplify our expressions we
introduce $\nu$ creation and annihilation operators $a_l^A$ and
$\bar{a}_l^A$, $l=1,\ldots, \nu$, and construct a Fock space
vector
$$
|A\rangle \equiv \prod_{l=1}^\nu\prod_{i_l=1}^{h_l}
a_l^{C_l^{i_l}}
A_{\phantom{1}}^{C^1_1,\ldots, C^{h_1}_1,\ldots,
C^1_\nu,\ldots, C^{h_{_\nu}}_\nu}|0\rangle\,,
\quad
[\bar{a}_i^A,\, a_j^B]=
\eta^{AB}\delta_{ij}\,,\quad
\bar{a}_n^A|0\rangle=0\,.
$$
For a realization of this kind, $M^{AB}$ has the form
$$
M^{AB}= \sum_{l=1}^\nu
(a_l^A\bar{a}_l^B-a_l^B\bar{a}_l^A)\,.
$$
Throughout of the paper, unless otherwise specified,
the indices $i,j,l,n$ run over $1,\ldots, \nu$. For these indices
we drop the summation over repeated indices.
Because the $A^{C({\bf h})}$ is associated
with ${\sf YT}({\bf h})$ then the $|A\rangle$ should
satisfy the constraints

\begin{equation}\label{ytfcon1}
(a_{ii}^{}-h_i)|A\rangle=0\,,\quad
a_{ij}^-|A\rangle=0\,,\quad
\varepsilon^{ij}_{} a_{ij}^{}|A\rangle=0\,,
\end{equation}
where in (\ref{ytfcon1}) and below we use the notation
\begin{equation}\label{fockv}
a_{ij}^{}\equiv  a_i^A \bar{a}_j^A\,,
\qquad
a_{ij}^-\equiv \bar{a}_i^A \bar{a}_j^A\,,
\qquad
a_{ij}^+\equiv a_i^A a_j^A\,,
\end{equation}
and $\varepsilon^{ij}=1(0)$ for $i<j(i\geq j)$.
The 1st equation in (\ref{ytfcon1}) tells us that
$a_i$ occurs $h_i$ times in $|A\rangle$.
Tracelessness of $A^{C({\bf h})}$
is reflected in the 2nd equation in (\ref{ytfcon1}).
The 3rd equation in (\ref{ytfcon1}) implies that the generic
tensor field (\ref{generic}) is antisymmetric with respect to
indices in columns.  As a result the $|A\rangle$ is obtainable
from YT by making use of the following
symmetrization rule:  (i) first we perform alternating with
respect to indices in all columns, (ii) then we perform
symmetrization with respect to indices in all rows.  Note that
usual one uses the symmetrization rule when first one performs
(ii) and then (i).  Such kind of $|A\rangle$ could be described
by using generic tensor field (\ref{generic}) which is symmetric
with respect to indices in columns and by using anticommuting
oscillators in place of commuting ones.

Because, by assumption, the $|A\rangle$ is a carrier for
$D(E_0,{\bf h})$ it should satisfy the equation
$$
(Q-\langle Q\rangle)|A\rangle=0\,,
$$
where  $Q$ is the second
order Casimir operator of the $so(d-1,2)$ algebra
while $\langle Q\rangle$ is its eigenvalue for
$D(E_0, {\bf h})$
$$
Q\equiv\frac{1}{2}J^{AB}J^{AB}\,,
\qquad
\langle Q\rangle =-E_0(E_0+1-d)
-\sum_{l=1}^\nu h_l (h_l-2l +d-1)\,.
$$
In addition we impose on $|A\rangle$ the following subsidiary
constraints

\begin{eqnarray}
\label{div1}
&&\bar{\nabla}_n|A\rangle=0 \qquad (\hbox{divergenelessness})\,,
\\
&&
\label{trans1}
\bar{y}_n |A\rangle=0
\qquad(\hbox{transversality})\,.
\end{eqnarray}
Here and below we use the notation
$$
\bar{\nabla}_n\equiv  \nabla^A\bar{\bf a}_n^A\,,
\quad
\nabla_n\equiv {\bf a}_n^A \nabla^A\,,
\quad
\bar{y}_n\equiv \bar{a}_n^A y^A\,,
\quad
y_n\equiv a_n^A y^A\,,
$$
$$
{\bf a}_n^A\equiv \theta^{AB}a_n^B\,,
\quad
\bar{\bf a}_n^A\equiv \theta^{AB}\bar{a}_n^B\,.
$$
The constraint (\ref{div1}) is a $SO(d-1,2)$ covariant analog
of usual divergenelessness condition
$\partial_\mu A^{\mu\ldots}=0$. The $SO(d-2,1)$ tensor
decomposes into the same rank tensor of Lorentz subgroup
$SO(d-1,1)$ and a lower rank tensor of $SO(d-1,1)$. The
constraint (\ref{trans1}) implies that the lower rank tensor is
set to zero. In other words we use a $SO(d-1,2)$
tensor which is irreducible when reducing to Lorentz subgroup.
Taking into account the transversality
(\ref{trans1}) and the relations
\begin{equation}\label{oscdecom}
a_{ij}^{}={\bf a}_{ij}-y_i\bar{y}_j\,,\qquad
a_{ij}^-={\bf a}_{ij}^- -\bar{y}_i\bar{y}_j,
\end{equation}
we transform the constraints
(\ref{ytfcon1}) to form which is more convenient in practical
calculations
\begin{equation}\label{ytfcon2}
({\bf a}_{ii}-h_{i})|A\rangle=0\,,\quad
{\bf a}_{ij}^-|A\rangle=0\,,\quad
\varepsilon^{ij}_{}{\bf a}_{ij}|A\rangle=0\,.
\end{equation}
Making use of constraints above the equations of motion
may be simplified. To this end we rewrite the $Q$ as follows
$$
Q=-\nabla^2+M^{AB}L^{AB}+\frac{1}{2}M^{AB}M^{AB}\,,
\qquad
\nabla^2\equiv \nabla^A\nabla^A\,,
$$
use then the relations
$$
M^{AB}L^{AB}|A\rangle=2\sum_{l=1}^\nu
h_l|A\rangle\,,
\qquad
M^{AB}M^{AB}|A\rangle
=-2\sum_{l=1}^\nu h_l(h_l-2l+d+1)|A\rangle
$$
and get the desired form of equations of motion
\begin{equation}\label{eqmot2}
(\nabla^2-m^2)|A\rangle=0\,,\qquad
m^2\equiv E_0(E_0+1-d)\,.
\end{equation}
To define $E_0$ corresponding to massless representations we
should construct gauge transformations and choose such the
$E_0$ that the equations (\ref{eqmot2}) to be invariant with
respect to gauge transformations.
In order to formulate gauge transformations we use the
gauge parameters fields whose spacetime indices correspond to
the $YT$ which one can make by removing one box from the ${\sf
YT}({\bf h})$.  The most general gauge
transformations we start with are
\begin{equation}\label{gaugetr1}
\delta_{(n)}|A\rangle\sim \nabla_n |\Lambda_n\rangle+
y_n |R_n\rangle\,,
\end{equation}
where the gauge parameters fields $|\Lambda_n\rangle$ and
$|R_n\rangle$ are associated with ${\sf YT}({\bf h}_{(n)})$,
and $i$-th component of the ${\bf h}_{(n)}$ is equal to
$h_{i(n)}=h_i-\delta_{in}$.  The ${\sf YT}({\bf h}_{(n)})$ is
obtained by removing one box from $n$-th row of the
${\sf YT}({\bf h})$.  We assume that only those
$|\Lambda_n\rangle$ and $|R_n\rangle$ are non-zero whose
${\bf h}_{(n)}$ satisfy the inequalities
\begin{equation}\label{gpwcon}
h_{1(n)}\geq\ldots \geq h_{\nu(n)}\geq 0\,.
\end{equation}
Given the ${\bf h}$, the set of those $n$ whose
${\bf h}_{(n)}$ satisfy (\ref{gpwcon}) will be referred to as
${\sf S({\bf h})}$.
We impose on
the  $|\Lambda_n\rangle$, $|R_n\rangle$ and the
constraints similar to those for $|A\rangle$
\begin{equation}\label{gpcon1}
\bar{\nabla}_i|\Lambda_n\rangle=0\,,
\qquad
\bar{y}_i|\Lambda_n\rangle=0
\end{equation}
and constraints obtained from (\ref{gpcon1}) by replacing
$\Lambda\rightarrow R$.
Since $|\Lambda_n\rangle$, $|R_n\rangle$ correspond
to ${\sf YT}({\bf h}_{(n)})$, they satisfy the constraints
\begin{equation}\label{ytgpcon1}
(a_{ii}^{}-h_{i(n)}^{})|\Lambda_n\rangle=0\,,\quad
a_{ij}^-|\Lambda_n\rangle=0\,,\quad
\varepsilon^{ij}_{} a_{ij}^{}|\Lambda_n\rangle=0\,,
\end{equation}
and those which are obtainable from (\ref{ytgpcon1})
by replacing $\Lambda\rightarrow R$. In practical calculation it
is convenient to rewrite (\ref{ytgpcon1}) in the form
\begin{equation}\label{ytgpcon2}
({\bf a}_{ii}^{}-h_{i(n)}^{})|\Lambda_n\rangle=0\,,\quad
{\bf a}_{ij}^-|\Lambda_n\rangle=0\,, \quad
\varepsilon^{ij}_{}{\bf a}_{ij}^{}|\Lambda_n\rangle=0\,.
\end{equation}
which can be obtained by using the constraints (\ref{gpcon1}).

It turns out that the invariance requirement of constraints
(\ref{trans1}),(\ref{ytfcon2})
with respect to gauge transformations fixes the form of gauge
transformations
\begin{equation}\label{gaugetr3}
\delta_{(n)}|A\rangle
={\cal D}_n|\Lambda_n\rangle\,,
\qquad
{\cal D}_n\equiv
\sum_{j=0}^{n-1}(-)^j\!\!\!\!\sum_{l_1,\ldots,l_{j+1}=1}^n
\delta_{nl_{j+1}} \prod_{i=1}^j
\frac{\varepsilon^{l_i l_{i+1}}}{\lambda_{l_i n}}
{\bf a}_{l_{i+1}l_i}^{} D_{l_1} \,,
\end{equation}
$$
D_n\equiv \nabla_n+\sum_{l=1}^\nu (-y_l^{}{\bf a}_{nl}^{}
+{\bf a}^+_{nl}\bar{y}_l^{})\,,
\qquad
\lambda_{ln}^{}
\equiv h_l^{} - h_n^{} +n-l +1\,.
$$
In (\ref{gaugetr3}) the ${\bf a}_{l_{i+1}l_i}$ are ordered as
follows:
${\bf a}_{l_{j+1}l_j}^{}\ldots {\bf a}_{l_2l_1}^{}\,.$
Then from the invariance requirement of (\ref{div1}) with
respect to (\ref{gaugetr3}), i.e.
$\bar{\nabla}_n\delta_{(n)}|A\rangle=0$, we find the equation of
motion for gauge parameter field
\begin{equation}\label{gpfeqmot1}
(\nabla^2-(h_n-n)(h_n-n-1+d))|\Lambda_n\rangle=0\,.
\end{equation}
Finally from the invariance requirement of equation of motion
(\ref{eqmot2}) with respect to gauge transformations
(\ref{gaugetr3}), i.e.  $(\nabla^2-m^2)\delta_{(n)}|A\rangle=0$,
we get the equation for $E_0$
\begin{equation}\label{quadeq}
E_0(E_0+1-d)=(h_n-n-1)(h_n-n-2+d)\,,
\quad n\in {\sf S({\bf h})}\,.
\end{equation}
Note that in deriving (\ref{quadeq}) the equations of
motion for gauge parameter has been used. Solutions to the
quadratic equation for $E_0$ (\ref{quadeq}) read:
\begin{equation}\label{posenval}
E_{0(n)}^{(1)}=h_n-n-2+d\,,
\qquad
E_{0(n)}^{(2)}=n+1-h_n\,.
\end{equation}
As seen from (\ref{posenval})
there exists an arbitrariness of $E_0$
parametrized by subscript $n$ which labels gauge
transformations and by superscripts $(1),(2)$ which label two
solutions of equation (\ref{quadeq}).
Because the values of $E_0$
have been derived by exploiting gauge invariance
we can conclude that the gauge invariance by itself does not
uniquely determine  the physical relevant value of $E_0$.  To
choose physical relevant value of $E_0$ we exploit the
unitarity condition, that is: 1) hermiticity $({\rm
i}J^{AB})^\dagger={\rm i}J^{AB}$; 2) the positive norm
requirement.
For details of resulting procedure we refer to (\cite{Metsit2})
and now let us formulate the result.

Given ${\sf YT}({\bf h})$ let $k$, $k=1\ldots\nu$,
indicates maximal number of upper rows which have the same
number of boxes. We call such Young tableaux the level-$k$
${\sf YT}({\bf h})$.  For the case of level-$k$ Young
tableaux the inequalities (\ref{mineq1}) can be rewritten as
\begin{equation}\label{mineq2}
h^{}_1=\ldots =h_k^{} > h_{k+1}^{} \ge
h^{}_{k+2}\ge \ldots \ge h_\nu^{}\geq 0.
\end{equation}
Then making use of unitarity condition one proves
(\cite{Metsit2}) that for the level-$k$ Young tableaux the
$E_0$ should satisfy the inequality
\footnote{This bound for $d=4$ has been found by
\cite{evans}. For the case $d=5$ see \cite{Mack} and references
therein.  Note that to extend (\ref{enbound}) to
odd $d$ we should simply replace $h_k\rightarrow |h_k|$. In view
that for odd $d$ all $h_i\ge 0$ with exception of
$h_{(d-1)/2}{\,=\kern -0.9em /\,\,}0$ our result are valid also
for odd $d$ when $h_{(d-1)/2}=0$. At present time field
theoretical description of representations for
arbitrary $h_{(d-1)/2}$ is absent.}
\begin{equation}\label{enbound}
E_0\ge h_k^{}-k-2+d\,.
\end{equation}
Comparing (\ref{posenval}) with (\ref{enbound})  we
conclude that only $E_{0(n=k)}^{(1)}$ satisfies the unitarity
condition.  Thus anti-de Sitter bosonic massless particles
described by level-$k$ ${\sf YT}({\bf h})$ takes lowest
value of energy equal to
\begin{equation}\label{envaltr}
E_0=h_k-k-2+d\,.
\end{equation}
Note that it is gauge transformation with $n=k$
(\ref{gaugetr3}) that leads to relevant $E_0$, i.e.
given level-$k$ ${\sf YT}({\bf h})$ only the gauge
transformation $\delta_{(k)}$ respects the unitarity.
Therefore only the $\delta_{(k)}$ will be used in what follows.
From now on we use letter $k$ to indicate level
of ${\sf YT}({\bf h})$. Thus the final form of gauge
transformation is
\begin{equation}\label{gaugetr4}
\delta_{(k)}|A\rangle
=\sum_{j=0}^{k-1}(-)^j\!\!\!\!\sum_{l_1,\ldots,l_{j+1}=1}^k
\delta_{kl_{j+1}}
\prod_{i=1}^j
\frac{\varepsilon^{l_i l_{i+1}}
{\bf a}_{l_{i+1}l_i}^{}}{k+1-l_i}
D_{l_1}|\Lambda_k\rangle\,.
\end{equation}
As an illustration of (\ref{gaugetr4}) we write down
$\delta_{(k)}$ for $k=1,2,3$:
$$
\delta_{(1)}|A\rangle=D_1^{}|\Lambda_1\rangle\,,
\qquad
\delta_{(2)}|A\rangle
=(D_2^{}-\frac{1}{2}{\bf a}_{21}^{}D_1^{})|\Lambda_2\rangle\,,
$$
$$
\delta_{(3)}|A\rangle
=(D_3
-\frac{1}{2}{\bf a}_{32}^{}D_2^{}
-\frac{1}{3}{\bf a}_{31}^{}D_1^{}
+\frac{1}{6}{\bf a}_{32}^{}{\bf a}_{21}^{}
D_1^{})|\Lambda_3\rangle\,.
$$
Note that from the
equation for gauge parameter field (\ref{gpfeqmot1}) we get the
following lowest energy value for $\Lambda_k$:
$E_0^\Lambda=E_0+1$, i.e.
\begin{equation}\label{gpenval}
E_0^\Lambda=h_k-k-1+d\,.
\end{equation}
With the values for $E_0^\Lambda$ at hand we are ready to
provide an answer to the question: do the gauge parameter
fields meet the masslessness criteria?
Because the inter-relation between of spin ${\bf h}$ and
energy value $E_0$ for massless field is given by
(\ref{envaltr}) we should express the $E_0^\Lambda$ in terms
of ${\bf h}^\Lambda$ and $k^\Lambda$, where $k^\Lambda$ is a
level of ${\sf YT}({\bf h}^\Lambda)$. Due to relations
$k^\Lambda=k-1$, $h^\Lambda_{k^\Lambda}=h_k-\delta_{k1}$ we
cast (\ref{gpenval}) to
$$
E_0^\Lambda=h_{k^\Lambda}^\Lambda-k^\Lambda-2+d+\delta_{k1}\,.
$$
Comparing this relation with (\ref{envaltr}) we conclude that
only for $k>1$ the gauge parameters are massless fields while
for $k=1$ they are massive fields.
Thus we have constructed equations of motion (\ref{eqmot2})
which respect gauge transformations (\ref{gaugetr4}), where the
gauge parameter fields $\Lambda_k$ satisfy the constraints
(\ref{gpcon1}), (\ref{ytgpcon2}) and equations of motion
(\ref{gpfeqmot1}).  The relevant $E_0$ and
$E_0^\Lambda$ are given by (\ref{envaltr}) and (\ref{gpenval}).

All things above have been done in $SO(d-1,2)$ covariant form.
Because sometimes a formulation in terms of intrinsic
coordinates is preferable let us transform our result to such
the coordinates.
Let $x^\mu$, $\mu=0,1,\ldots, d-1$ be the intrinsic
coordinates in AdS spacetime and let $y^A(x)$ be
imbedding map, where $y^A(x)$ satisfy (\ref{hyperbol}).  The
relationship between $SO(d-1,2)$ tensor field
$A^{C_1\ldots}$ and the usual tensor field $A^{\mu\ldots}$
is given by
$$
A^{\mu_1\ldots}(x)=y_{C_1}^{\mu_1}\ldots A^{C_1\ldots}(y)\,,
\qquad
y^\mu_C\equiv g^{\mu\nu}\partial_\nu y_C^{\vphantom{5pt}}\,,
$$
where the intrinsic geometry metric tensor is given by
$g_{\mu\nu}=\partial_\mu y^A \partial_\nu y^A$ while its
inverse is $g^{\mu\nu}=\nabla^A x^\mu \nabla^A x^\nu$.
The $x^\mu=x^\mu(y)$ is a certain representation of intrinsic
coordinates. There are useful relations
$$
\theta^{AB}=g^{\mu\nu}\partial_\mu y^A \partial_\nu y^B\,,
\qquad
\partial_\mu y^A \nabla^A x^\nu =\delta_\mu^\nu\,,
$$
$$
\nabla^A x^\mu =g^{\mu\nu} \partial_\nu y^A\,,
\quad
\nabla^2 x^\mu =-\Gamma_{\rho\sigma}^\mu g^{\rho\sigma}
\,,
\quad
D_\mu y_\nu^A=g_{\mu\nu} y^A\,.
$$
With these relation at hand and with the help of relations
$$
y_{C_1}^{\mu_1}\ldots y_{C_s}^{\mu_s}
\nabla^2 A^{C_1\ldots C_s}
=D^2 A^{\mu_1\ldots \mu_s}
+s A^{\mu_1\ldots\mu_s}\,,
\qquad
D^2 A^{C_1\ldots }=\nabla^2 A^{C_1\ldots}
$$
where
$D^2\equiv D_\mu D^\mu$,
$D_\mu=\partial_\mu+\Gamma_{\mu\cdot}^\cdot$,
we can immediately transform equation of motion (\ref{eqmot2}) to
the desired form
\begin{equation}\label{eqmot4}
(D^2-(h_k-k-1)(h_k-k-2+d)
+\sum_{l=1}^\nu h_l)A^{\mu_1\ldots}=0\,.
\end{equation}
It turns out that in order to write gauge transformation it is
convenient to transform spacetime tensors into the tangent
space tensors
$A^{a_1\ldots}\equiv e_{\mu_1}^{a_1}\ldots A^{\mu_1\ldots }$
where the $e_\mu^a$ is a einbein of AdS geometry,
introduce new
creation and annihilation operators $a_l^a$ and $\bar{a}_l^a$,
$l=1,\ldots\nu$, $a=0,1,\ldots,d-1$, and construct Fock space
vector
$$
|a\rangle=a^{a_1}\ldots A^{a_1\ldots}|0\rangle\,,
\qquad
[\bar{a}_i^a,a_j^b]=\delta_{ij}\eta^{ab},
\qquad
\eta^{ab}=(-,+,\ldots,+)\,.
$$
Now the equation, constraints and gauge transformation take the
form
\begin{equation}\label{fineq}
(D_L^2-(h_k-k-1)(h_k-k-2+d)
+\sum_{l=1}^\nu h_l)|a\rangle=0\,,
\end{equation}
$$
\delta_{(k)}|a\rangle
=\sum_{j=0}^{k-1}(-)^j\!\!\!\!\sum_{l_1,\ldots,l_{j+1}=1}^k
\delta_{kl_{j+1}}
\prod_{i=1}^j
\frac{\varepsilon^{l_i l_{i+1}}
(a_{l_{i+1}}^{} \bar{a}_{l_i}^{})}{k+1-l_i}
a^b_{l_1}
e_b^\mu D_{\mu L}|\lambda_k\rangle\,,
$$

\begin{equation}\label{fincon}
\bar{a}_i^be_b^\mu D_{\mu L}|a\rangle=0\,,
\quad
(a_i^b\bar{a}_i^b-h_i)|a\rangle=0\,,
\quad
\bar{a}_i^b\bar{a}_i^b|a\rangle=0\,,
\quad
\varepsilon^{ij}a_i^b\bar{a}_j^b|a\rangle=0\,,
\end{equation}
$$
D_{\mu L}\equiv \partial_\mu+\frac{1}{2}\omega_\mu^{ab}M^{ab}\,,
\qquad
M^{ab}\equiv\sum_{l=1}^\nu (a_l^a\bar{a}_l^b-a_l^b\bar{a}_l^a)\,.
$$
The $\omega_\mu^{ab}$ is a Lorentz connection of AdS spacetime.
The equation and constraints for the gauge parameter
field $\lambda_k$ are obtainable from (\ref{fineq}) and
(\ref{fincon}) by making there the substitutions
$|a\rangle\rightarrow |\lambda_k\rangle$, $h_i\rightarrow
h_{i(k)}$ and $k\rightarrow k-2$.  In order to demonstrate how
our results are working let us consider some particular cases.

{\it Totally antisymmetric fields}.
In this case ${\bf h}=(1,\ldots,1,\ldots,0)$ where unit occurs
s-times in this sequence. Therefore we have
$k=s$ and $h_k=\varepsilon^{ks+1}$.
For this case $E_0=d-1-s$ and
we get the equations
$$
(D^2+s(d-s))A^{\mu_1\ldots\mu_s}=0\,,
$$
where the relevant constraint is
$D_\mu A^{\mu\mu_2\ldots \mu_s}=0$.
By making use of this gauge the equations above can be easily
derived from well known equations
$$ D_\mu
F^{\mu\mu_1\ldots\mu_s}=0\,, \qquad F_{\mu_1\ldots \mu_n} \equiv
n\partial_{[\mu_1} A_{\mu_2\ldots \mu_n]}\,.
$$

{\it Totally symmetric fields.}
In this case ${\bf h}=(s,0\ldots,0)$.
Therefore we have $k=1$ and $h_k=s\varepsilon^{k2}$,
the $E_0$ is given by $E_0=s+d-3$
and we get the equations
\begin{equation}\label{totsym}
(D^2-s^2+(6-d)s+2d-6)A^{\mu_1\ldots\mu_s}=0\,,
\end{equation}
where the relevant constraints are
\begin{equation}\label{simg}
A_\mu{}^{\mu\mu_3\ldots\mu_s}=0\,,
\qquad
D_\mu A^{\mu\mu_2\ldots\mu_s}=0\,.
\end{equation}
For $d=4$ these equations can be obtained from those
discovered in (\cite{F2}) by making use of the gauge (\ref{simg}).
The {\it graviton} is a particular case when $s=2$. From
(\ref{totsym}) we get the equation
$(D^2+2)h_{\mu\nu}=0$ which should be supplemented by constraints
like (\ref{simg}). This equation coincides with that
obtained from Einstein equation for excitation of metric
tensor
$$ R_{\mu\nu}=-(d-1)G_{\mu\nu}\,, \quad
G_{\mu\nu}=g_{\mu\nu}+h_{\mu\nu}\,,
\quad
R_{\mu\nu\rho\sigma}(g)
=-(g_{\mu\rho}g_{\nu\sigma}
-g_{\mu\sigma}g_{\nu\rho})
$$
where $g_{\mu\nu}$ is a metric tensor of AdS
geometry. Note that the value $s=2$ is the only when
the dependence on $d$ in (\ref{totsym}) is cancelled.  Thus we
have demonstrated that our results cover all previously known
particular cases and solve problem for arbitrary spin ${\bf h}$
massless bosonic fields  in $d$ - dimensional AdS spacetime.

In conclusion let us discuss masslessness in $d$ - dimensional AdS
spacetime by using the requirement of conformal invariance.
By conformal invariant representations we will understand those
irreducible representations of anti-de Sitter group that
can be realized as irreducible representations of the conformal
group $SO(d,2)$. It turns out (for details see \cite{Metsit5}))
that this requirement leads to representations whose $h_i$ satisy
the constaints $h_1=\ldots=h_\nu\equiv h$, while $E_0=h+\nu$.
These $E_0$ and $h_i$ are in accordance with
(\ref{envaltr}), i.e. conformal invariance
respects the gauge invariance, but because of constraints
above-mentioned
the conformal representations constitute only a subset of all
massless states for $d>4$, i.e.  conformal group for $d>4$
cannot be used for defining all massless representations.
In this respects the situation in AdS spacetime
(\cite{Metsit5}) is similar to that in Minkowski spacetime
(\cite{sig}).

{\it Acknowledgements}.
This work was supported in part by the RFBR, Grant
96-02-17314a and by RFBR Grant for Leading Scientific
Schools N 96-15-96463.

\end{document}